\documentclass[a4paper,11pt]{article}
\usepackage{pos}

\def\be{\begin{equation}}
\def\ee{\end{equation}}
\def\bse{\begin{subequations}}
\def\ese{\end{subequations}}
\def\bal{\begin{align}}
\def\ealn{\end{align}}

\def\bs{\boldsymbol}
\def\bp{\begin{pmatrix}}
\def\ep{\end{pmatrix}}

\def\bse{\begin{subequations}}
\def\ese{\end{subequations}}
\def\bal{\begin{align}}
\def\ealn{\end{align}}

\def\bs{\boldsymbol}

\title{ Quantum matrix geometry in the lowest Landau level and higher Landau levels}

\author{Kazuki Hasebe}

\affiliation{National Institute of Technology, Sendai College,  \\
  Ayashi, Sendai, 989-3128, Japan}

\emailAdd{khasebe@sendai-nct.ac.jp}

\abstract{ 
One of the most celebrated works of Professor Madore is the introduction of fuzzy sphere. 
I briefly review  how the fuzzy two-sphere and its higher dimensional cousins   are realized  
in the (spherical) Landau models in non-Abelian monopole backgrounds. For extracting  quantum geometry from the Landau models, we  evaluate the matrix elements of the coordinates of spheres in the lowest and higher Landau levels. 
For the lowest Landau level, the matrix geometry is identified as the geometry of  fuzzy sphere. 
Meanwhile for the higher Landau levels,   the obtained quantum geometry turns out to be a nested matrix geometry with no classical counterpart. 
There exists a hierarchical structure between the fuzzy geometries and the monopoles in different dimensions.  
That dimensional hierarchy signifies a Landau model counterpart of 
the dimensional ladder of quantum anomaly. 
    }

\FullConference{%
  Corfu Summer Institute 2021 "Workshops on Quantum Geometry, Field Theory and Gravity"\\
  20 September - 27 September 2021\\
  Corfu, Greece
}

\begin{document}
\maketitle

\section{Introduction}

The non-commutative geometry is a promising mathematical framework for describing a quantized space-time. 
Typical and well-studied examples of the non-commutative spaces are  non-commutative plane,  fuzzy sphere and fuzzy hyperboloid.  They exhibit  mathematically unique structures and also exemplify   solutions of the Matrix models of string theory. 

 Usually,  
 a non-commutative structure is postulated when defining  models such as in the non-commutative field theory, and their physical properties  are discussed withing the given framework.     
Here,  we will take an almost inverse approach: We will 
 introduce Landau models at first not assuming any non-commutative structure  and subsequently exploit the non-commutative structure from  the models.  
The Landau models are  simple models that describe quantum mechanics of electrons in  magnetic fields. 
In the planar Landau model,  the center-of-mass coordinates of electron are given by    
$X=x+i\frac{1}{B}(\partial_y+iA_y)$ and $Y=y-i\frac{1}{B}(\partial_x+iA_x)$  
that satisfy the Heisenberg-Weyl algebra:  
\be
[X, Y] =i\frac{1}{B}\boldsymbol{1}, ~~~~~[X, \boldsymbol{1}] =[Y, \boldsymbol{1}]=0.
\ee
The center-of-mass coordinates obey the non-commutative algebra due to the existence of the magnetic field and the electrons  behave as if they live on a non-commutative plane. As for  curved non-commutative spaces,  
the fuzzy sphere introduced by Madore \cite{madore1992}  (see  Refs.\cite{Berezin1975, Hoppe1982} also)  provides a typical example.    
The mathematics of fuzzy two-sphere  is very simple: 
The matrix coordinates satisfy the $SU(2)$ algebra and the sum of their squares should be   a constant (times an identity matrix): 
\be
[X_i, X_j] =i\epsilon_{ijk}X_k,~~~~~~\sum_{i=1}^3 X_iX_i =\text{const}\cdot \bs{1}. \label{fuzzytwosphere} 
\ee
After  a few years of the work of Madore, Grosse and his collaborators succeeded to generalize the concept of the fuzzy-sphere  in the four dimension  \cite{Grosse-Klimcik-Presnajder-1996}, which is now known as the fuzzy four-sphere.   The  fuzzy four-sphere was rediscovered in the context of string theory  and studied in detail  \cite{Castelino-Lee-Taylor-1997,  Ho-Ramgoolam-2002, Kimura2002, Kimura2003}. The  fuzzy four-sphere is defined as   
\be 
[X_a, X_b, X_c, X_d] =\epsilon_{abcde}X_e,~~~~~\sum_{a=1}^5 X_a X_a =\text{const}\cdot \bs{1}. \label{fuzzyfoursphere}
\ee 
Obviously,  (\ref{fuzzyfoursphere}) is a straightforward generalization of (\ref{fuzzytwosphere})  replacing the usual commutator with  the Nambu  four-bracket  
$[X_{\sigma_1}, X_{\sigma_2}, X_{\sigma_3}, X_{\sigma_4}] \equiv  \text{sgn}(\sigma) ~ X_{\sigma_1} X_{\sigma_2} X_{\sigma_3} X_{\sigma_4}$   
\cite{Nambu1973, CurtrightZachos2003, DeBellisSS2010}. 
Meanwhile, the fuzzy three-sphere  \cite{Guralnik&Ramgoolam2001, Ramgoolam2001, Ramgoolam2002, BasuHarvey2004, Jabbari2004, JabbariTorabian2005} is introduced as 
\be 
[\![X_\mu X_\nu, X_\rho]\!]=i\epsilon_{\mu\nu\rho\sigma}X_\sigma, ~~~~~~~\sum_{\mu=1}^4 X_\mu X_\mu =\text{const}\cdot \bs{1}.  \label{fuzzythreesphere}
\ee 
Note that, in (\ref{fuzzythreesphere}),  the ``three bracket''  $[\![X_\mu X_\nu, X_\rho]\!] \equiv [X_\mu, X_\nu, X_\rho, G_5]$ with  $G_5\equiv P_{R_+} -P_{R_-} =\begin{pmatrix} 
1 & 0 \\
0 & -1
\end{pmatrix}$ is used.   

In the following, I will discuss the Landau model realization of  the fuzzy pheres.\footnote{In the same spirit,  the fuzzy $\mathbb{C}P^n$ is analyzed in \cite{KarabaliNair2002}. See \cite{Karabali-Nair-Daemi-2004} as a review.}  Before going to details, I briefly  mention the underlying idea.  
Suppose that we would like to construct a fuzzy manifold whose classical geometry is a coset, $M \simeq G/H$. 
The corresponding fuzzy manifold will be made of ``quantum elements'' each of which occupies some quantum finite area on $M$. As $M$ returns to itself under the transformation $G$, the set of those quantum elements should return to itself under the  transformation.      We may adopt the states of an irreducible representation of $G$ as  such quantum elements, since irreducible representation denotes a closed set  under the group transformations.   
  Thus, the fuzzy manifold made of the states of the irreducible representation is necessarily symmetric under the  transformation $G$.  
The stabilizer group $H$ of $M$ will be translated as the gauge group in the quantum mechanical side. 
In short, in passing from the classical geometry $M \simeq G/H$ to its fuzzy version $M_F$, we need to utilize the irreducible representations of  the  global symmetry group $G$ with the gauge symmetry $H$.   
Such a quantum mechanical model is nothing but the Landau model constructed on $M$ with the gauge symmetry $H$.

\section{2D Landau model and  fuzzy two-sphere}

To fuzzify a two-sphere $S^2\simeq SO(3)/SO(2)$,  we consider the 
Landau model with  the $SO(2)\simeq U(1)$ gauge symmetry on $S^2$. 
 The Landau model Hamiltonian is given by \cite{Wu-Yang-1976, Haldane-1983} 
\be
H=-\frac{1}{2M}\sum_{i=1}^3(\partial_i+iA_i)^2|_{r=\text{const}} 
\ee
where $A_i$ denote the Dirac's monopole gauge field  \cite{Dirac-1931}
\be
A_{\mu=1,2} =-\frac{I}{2r(r+x_3)}\epsilon_{\mu\nu}x_\nu, ~~~A_3 =0. 
\ee
The corresponding Landau levels are 
$E_n =\frac{1}{2M}(I(n+\frac{1}{2}) +n(n+1))$ with $n=0,1,2,\cdots$ being the Landau level index,  
and the Landau level degeneracy is counted as  
$d_n =2n+I+1$. 
The $n$th Landau level eigenstates are known as  the monopole harmonics    
$Y_{l=n+\frac{I}{2}, m}^{(I/2)}(\theta,\phi)$ \cite{Wu-Yang-1976} which  
constitute the $SU(2)$ irreducible representation of the spin index $l=n+\frac{I}{2}$.

\begin{figure}[tbph]
\center
\hspace{-0.8cm}
\includegraphics*[width=70mm]{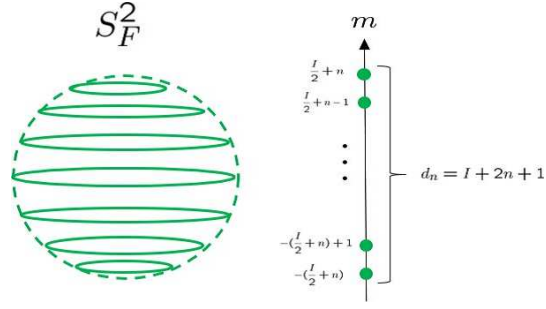}
\caption{The  $n$th Landau level eigenstates (monopole harmonics) with the $SU(2)$ Casimir index $l=n+\frac{I}{2}$ constitute the fuzzy two-sphere $X_i^{(n)}$.    
}
\label{fuzzys2.fig}
\end{figure}

While there are several  methods for deriving the non-commutative geometry from the Landau model,  the  most straightforward  way is to evaluate the matrix elements of the coordinates in each Landau level: 
\be
(X_i^{(n)})_{mm'} =\langle Y_{l, m}|x_i|Y_{l, m'}\rangle|_{l=n+\frac{I}{2}}. \label{xnmmsu2}
\ee
Using the formulas of the monopole harmonics, we can explicitly evaluate  (\ref{xnmmsu2}) as \cite{Hasebe-2015}  
\be
X_i^{(n)} =\frac{2I}{(I+2n)(I+2n+2)}S_i^{(l=n+\frac{I}{2})},  \label{ximat}
\ee
where $S^{(l)}$ denote the $SU(2)$ spin matrices with spin magnitude $l$. 
The $X_i^{(n)}$ (\ref{ximat}) obviously satisfy the algebra of the fuzzy two-sphere (\ref{fuzzytwosphere}):   
\be
[X_i^{(n)},X_j^{(n)}] =i\epsilon_{ijk} \frac{(I+2n)(I+2n+2)}{2I}X_k^{(n)},~~~~~\sum_{i=1}^3 X_i^{(n)}X_i^{(n)}  =\frac{I^2}{(I+2n)(I+2n+2)} \boldsymbol{1}_{2n+I+1}.
\ee
The $U(1)$ quantum number $m$ appears as the $2l+1$  diagonal components of 
$X_z^{(n)}~\propto~S_z^{(l=n+\frac{I}{2})}$ and specifies  the position of the latitudes    
(Fig.\ref{fuzzys2.fig}).    
While 
 the non-commutative structure of the lowest Landau level $(n=0)$ is usually focused in literature, the non-commutative geometry also appears  in the higher Landau levels $(n\neq 0)$ as shown by (\ref{ximat}).

\section{4D Landau model and  fuzzy four-sphere}

\begin{figure}[tbph]
\center
\hspace{-0.8cm}
\includegraphics*[width=60mm]{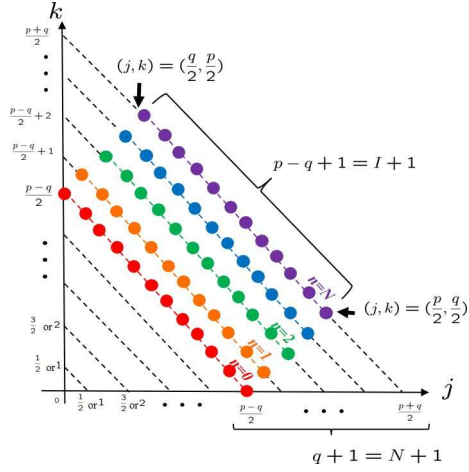}
\caption{The $SO(5)$ irreducible representation for $(p, q)=(I+N,N)$. The $SO(5)$ irreducible representation consists of the $N+1$ oblique lines on which  $SO(4)$ irreducible representations are aligned. (Taken from \cite{Hasebe-2020}.)}
\label{so5irr.fig}
\end{figure}

Next I discuss the fuzzy four-sphere geometry.   
For $S^4 \simeq SO(5)/SO(4)$,  we need to consider the Landau model  on $S^4$   in the $SO(4)$ monopole background. The $SO(4)$ group is a direct product of two $SU(2)$ groups, and then we will adopt one of them.\footnote{Recently, the author performed a full analysis for the $SO(4)$  case   \cite{Hasebe-2021}. }   
With the  $SU(2)$ monopole gauge field \cite{Yang-1978-1}
\be
A_{\mu=1,2,3,4} =-\frac{1}{2r(r+x_5)}\eta_{\mu\nu}^i x_{\nu}S_i^{(I/2)}, ~~~A_5 =0, ~~~~~~~(\eta_{\mu\nu}^i =\epsilon_{\mu\nu i4}+\delta_{\mu i}\delta_{\nu 4}-\delta_{\mu 4}\delta_{\nu i}), 
\ee
 the 4D Landau Hamiltonian is constructed as  \cite{Zhang-Hu-2001}. 
\be
H=-\frac{1}{2M}\sum_{a=1}^5{(\partial_a+iA_a)}^2 |_{r=\text{const.}}.
\ee
The Landau levels are  given by 
$E_N =\frac{1}{2M}(I(N+1) +N(N+3))$ $(N=0,1,2,\cdots)$. 
The corresponding  eigenstates are 
known as  the $SU(2)$ monopole harmonics $Y_{p, q;~ j, m_j; k, m_k}$  \cite{Yang-1978-2} with the $SO(5)$ Casimir indices 
\be
(p, q) =(N+I, N) ~~~(N=0,1,2,\cdots).
\ee
Note that the $SU(2)$ monopole harmonics carry the $SO(4)\simeq SU(2) \otimes SU(2)$ quantum numbers $(j,m_j;~k,m_k)$, which brings  
$D(N, I)  
\equiv \frac{1}{6}(N+I+2)(N+1)(2N+I+3)(I+1)$    
degeneracy to the $N$th Landau level, and   
the $N$th Landau level  eigenstates 
 consist  of $N+1$ sets of the  $I+1$   $SO(4)$ irreducible representations (see  Fig.\ref{so5irr.fig}). The  $SO(4)\simeq SU(2)\times SU(2)$ decomposition of the $SO(5)$ irreducible representation is given by 
\be
D(N, I) =\sum_{n=0}^N \sum_{j+k =n+\frac{I}{2}} (2j+1)(2k+1).  \label{decso5toso4}
\ee
With those Landau level eigenstates, we derive  the matrix coordinates 
\be
(X_a^{(N)})_{j, m_j; k, m_k}=\langle Y_{N+I, N; j, m_j; k, m_k}|x_a|Y_{N+I, N; j, m_j; k, m_k}\rangle, \label{foursmat}
\ee
which will become  $D(N, I)\times D(N, I)$ matrices.

\subsection{Matrix geometry from the lowest Landau level }

\begin{figure}[tbph]
\center
\hspace{-0.8cm}
\includegraphics*[width=90mm]{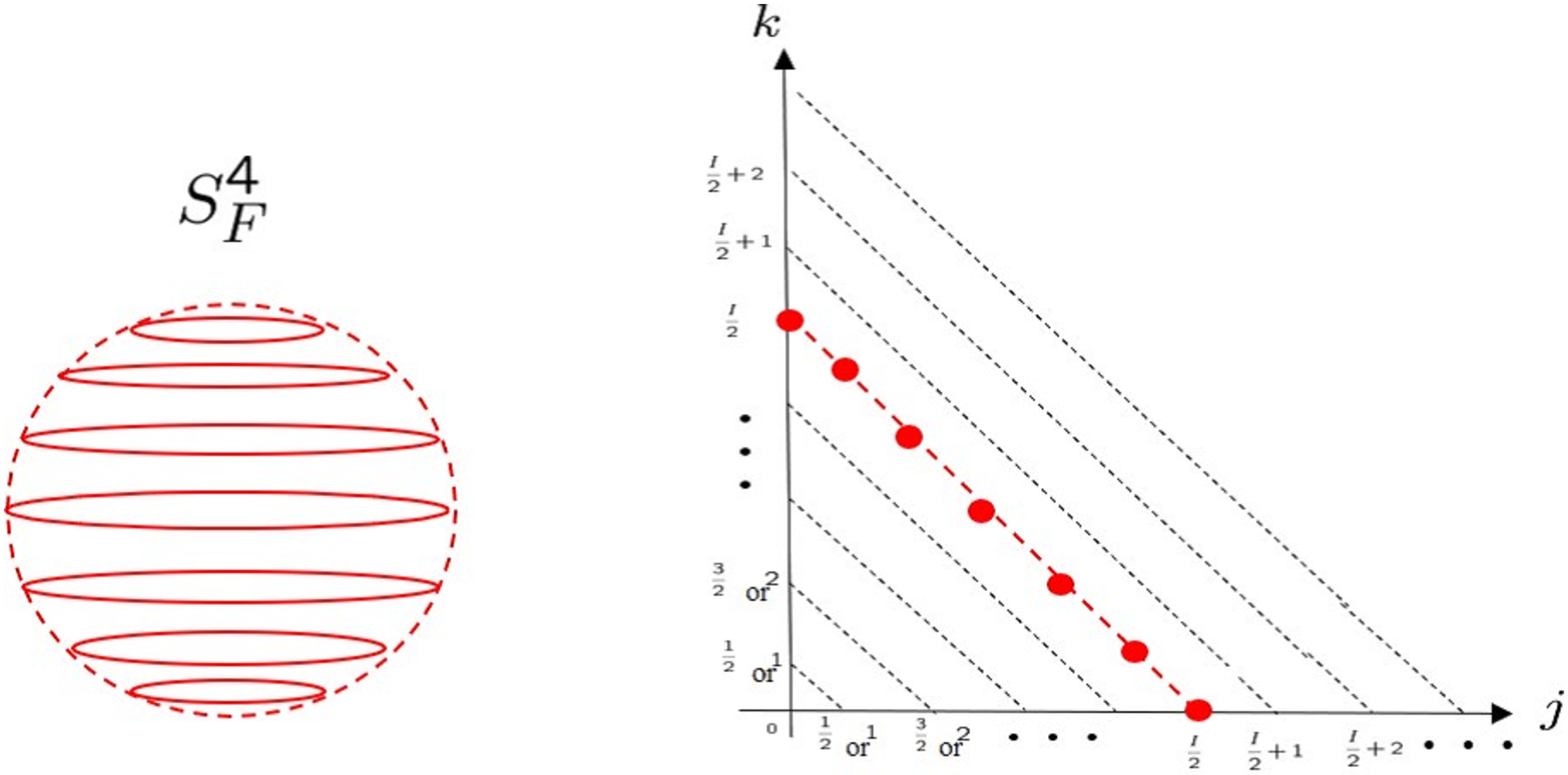}
\caption{The $S^3$-latitudes on the fuzzy four-sphere (left) correspond to the $SO(4)$ irreducible representations (right).}
\label{so5lll.fig}
\end{figure}

In the lowest Landau level $(N=0)$, the matrix coordinates (\ref{foursmat}) are obtained as \cite{Hasebe-2020} 
\be
X_a^{(0)} =\frac{1}{I+4}\Gamma_a,  \label{so5lllmat}
\ee
where $\Gamma_a$ denote $I$ tensor product of the $SO(5)$ gamma matrices.\footnote{The authors \cite{Ishiki-Matsumoto-Muraki-2018} derived the result  (\ref{so5lllmat}) in the context of  the Berezin-Toeplitz quantization.} 
The $X_a^{(0)}$ (\ref{so5lllmat}) actually satisfy  the relations  of the fuzzy four-sphere (\ref{fuzzyfoursphere}):   
\be
[X_a^{(0)}, X_b^{(0)},  X_c^{(0)}, X_d^{(0)}] = 8 \frac{(I+2)}{(I+4)^3} \epsilon_{abcde}X_e^{(0)},~~~~~\sum_{a=1}^5X_a^{(0)} X_a^{(0)}  =\frac{I}{I+4} \boldsymbol{1}. 
\ee
$X_5$ is a diagonal matrix whose diagonal elements represent  the difference between the two $SU(2)$ Casimir indices, $j-k$, and  specify  
the position of $S^3$ latitudes  [Fig.\ref{so5lll.fig}].  
Each of the $S^3$-latitudes accommodates the  degeneracy $(2j+1)(2k+1)$ due to the   $SO(4)\simeq SU(2)\otimes SU(2)$ internal  structure. 

\subsection{Nested matrix geometry from the higher Landau levels }

\begin{figure}[tbph]
\center
\includegraphics*[width=140mm]{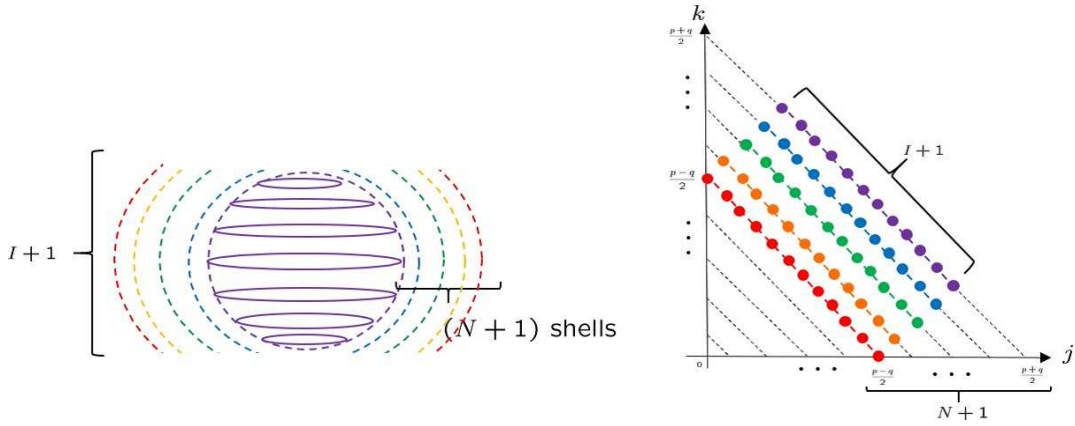}
\caption{ The $(N+1)$ fuzzy shells realize the nested geometry  of    the $N$th  Landau level. }
\label{nestedspheresfig}
\end{figure}

The decomposition rule (\ref{decso5toso4})   from $SO(5)$ to $SO(4)$  
 implies a nested structure in the corresponding  matrix geometry \cite{Hasebe-2020}: 
Each oblique line of the $SO(4)$ irreducible representations corresponds to a  ``fuzzy shell'' in the matrix geometry side, and  
the  $(N+1)$ sets of the $SO(4)$ irreducible representations constitute the nested structure of the $(N+1)$ shells   [Fig.\ref{nestedspheresfig}].  

\begin{figure}[tbph]
\center
\hspace{-0.8cm}
\includegraphics*[width=150mm]{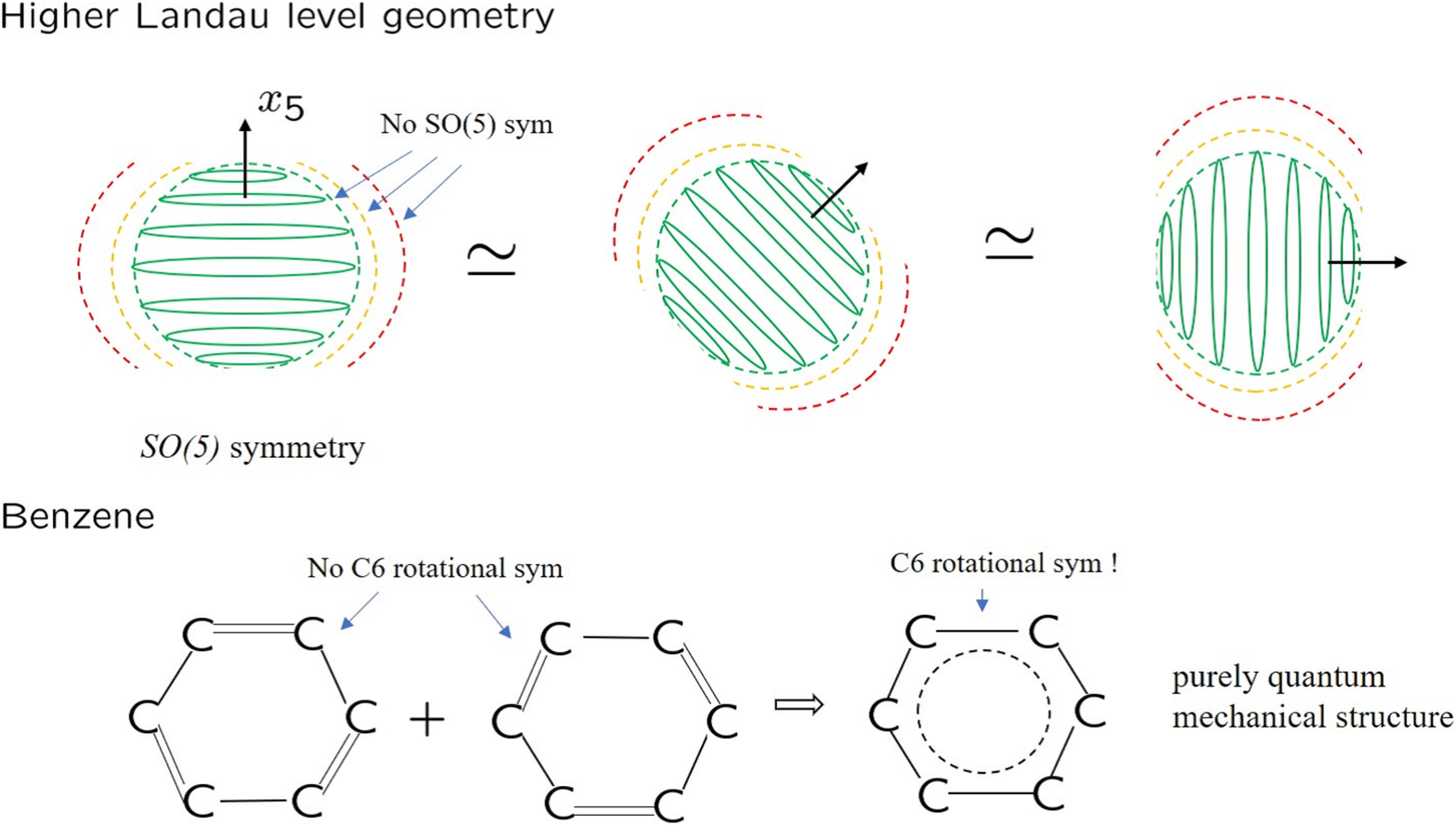}
\caption{ Analogies between the nested geometry of higher Landau level and the covalent bond  of benzene.   }
\label{benzene.fig}
\end{figure}

Each of the fuzzy shells  does not respect the $SO(5)$ symmetry, 
but the set of the fuzzy shells gives rise to the $SO(5)$ symmetric fuzzy manifold composed of  the  states of the $SO(5)$ irreducible representation.  
In Fig.\ref{nestedspheresfig},  the nested fuzzy manifold does not seem to have  the $SO(5)$ rotational symmetry, but  this is not the case. We have chosen the 5th axis as the quantization axis, but  
we can choose any axis in arbitrary direction [Fig.\ref{benzene.fig}]. In other words, that $SO(5)$ non-symmetric  picture of the nested fuzzy manifold in Fig.\ref{nestedspheresfig} is due to a  ``gauge artifact''.   
This situation is somewhat similar to the covalent bond of benzene [Fig.\ref{benzene.fig}]. The covalent band of benzene respects the C6 rotational symmetry,
 but either of the two Kekul\'{e} structures does not have the C6 rotational symmetry. Only the quantum composite of the two Kekul\'e structures  respects the C6 rotational symmetry as a whole. 
The covalent bond of benzene is a purely quantum mechanical structure with no classical counterpart.  
Back to the present fuzzy geometry, each of the fuzzy shells  does not respect the $SO(5)$ rotational symmetry, but their composite nested structure does the $SO(5)$ symmetry.  
In a similar sense of  benzene,  it is fair to say that the higher Landau level geometry realizes  a pure quantum geometry. 
See  \cite{Hasebe-2022-1} for details.  

\section{3D Landau model and  fuzzy three- and four-spheres}

\begin{figure}[tbph]
\center
\hspace{-0.8cm}
\includegraphics*[width=150mm]{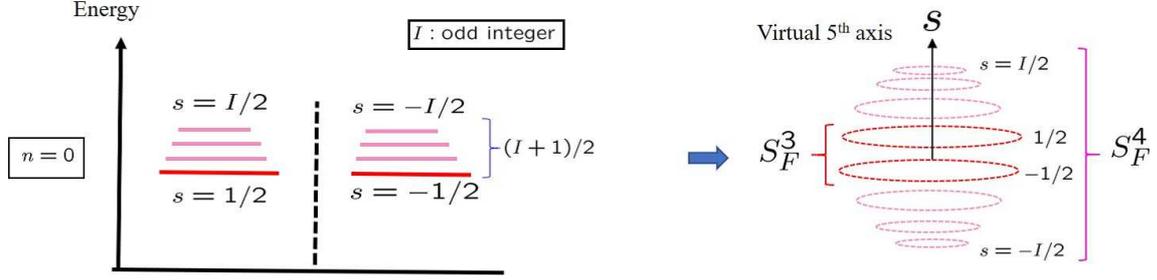}
\caption{ The lowest Landau level $(n=0)$ consists of $(I+1)$ sub-bands labeled by $s$ (left). 
 We align the fuzzy 3D  manifolds  along the virtual 5th direction to reproduce the fuzzy four-sphere geometry (right).  }
\label{so4l.fig}
\end{figure}

As the last concrete example, we will discuss the fuzzy three-sphere. 
For $S^3 \simeq SO(4)/SO(3)$, the corresponding Landau model  is constructed on $S^3$ in the $SO(3)\simeq SU(2)$ monopole background. That Landau model was first analyzed in \cite{Nair-Daemi-2004} and subsequently in  \cite{Hasebe-2014-2, Hasebe-2018}.  
The  $SU(2)$ monopole gauge field is given by 
\be
A_i =-\frac{I}{2r(r+x_4)}\epsilon_{ijk} x_jS_k^{(I/2)}, ~~~A_4 =0, 
\ee
and the  Landau Hamiltonian is 
\be
H=-\frac{1}{2M}\sum_{\mu=1}^4{(\partial_{\mu}}^2+iA_{\mu})|_{r=\text{const.}}.   
\ee
The Landau levels  are derived as  
$E_{n, s} =\frac{1}{2M}(I(n+\frac{1}{2}) +n(n+2)+s^2)$,  
which depend both on the Landau level index $n=0,1,2,\cdots$ and the sub-band index 
\be
s =\frac{I}{2}, ~\frac{I}{2}-1, ~\frac{I}{2}-2, ~\cdots, ~-\frac{I}{2}. 
\ee
The fuzzy three-sphere geometry is realized at the minimum energy level of $(n, s)=(0, 1/2 )$ and $(0, - 1/2)$. 
In the lowest Landau level $(n=0)$, there are $I+1$ sub-bands [Fig.\ref{so4l.fig}] labeled by  $s$, and the matrix coordinates of $S^3$ are obtained as 
\be
X_{\mu=1,2,3,4} =\frac{1}{I+3}\Gamma_{\mu}, 
\ee
where $\Gamma_{\mu=1,2,3,4}$ are  the four matrices of  (\ref{so5lllmat}). 
Aligning the fuzzy geometries in the sub-bands along the virtual 5th direction of  $s$,  
we can reproduce the fuzzy four-sphere geometry  [Fig.\ref{so4l.fig}] \cite{Hasebe-2014-2, Hasebe-2018}. 
Thus interestingly, the 3D Landau model  ``knows'' the geometry of the one dimension higher Landau model. 
Such a hierarchical structure has been observed  in the context of the fuzzy geometry  
\cite{Jabbari2004,  JabbariTorabian2005, Guralnik&Ramgoolam2001, Ramgoolam2001, Ramgoolam2002}, and  the present Landau models are the physical models that nicely realize the structure.

\section{Dimensional hierarchy }

\begin{figure}[tbph]
\hspace{-0.3cm}
\includegraphics*[width=150mm]{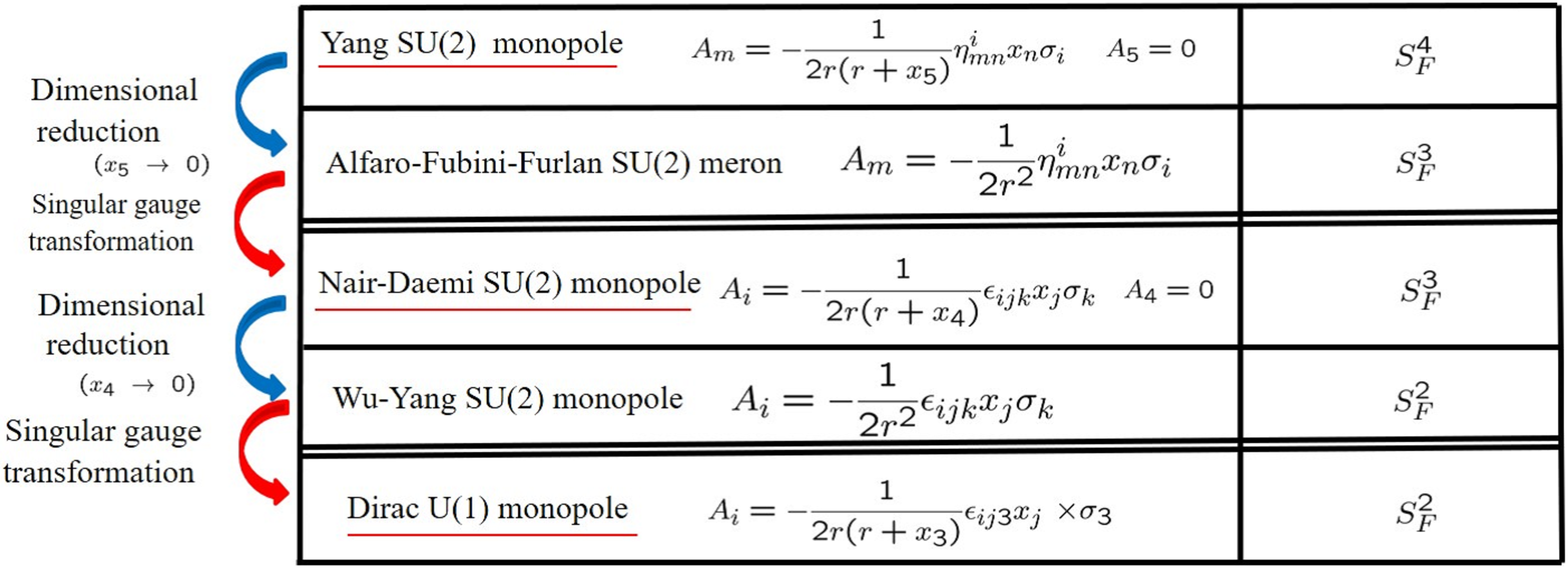}
\caption{The dimensional hierarchy of the monopole gauge fields. Starting from Yang's $SU(2)$ monopole in 5D, we reach Dirac's $U(1)$ monopole in 3D by applying the dimensional reductions and the singular gauge transformations.   (Taken from \cite{Hasebe-2020}.)   }
\label{gdim.fig}
\end{figure}

We can  find 
  a  similar hierarchical structure with respect to the monopole gauge fields  \cite{Hasebe-2020} [Fig.\ref{gdim.fig}]. 
One may wonder whether such a hierarchical relation is  extended in even higher dimensions.  Actually, it is straightforward to generalize  the set-up of the  Landau model in any dimensions, and we can show that the dimensional  hierarchy  ranges in all dimensions  \cite{Hasebe-Kimura-2003, Hasebe-2014-1, Hasebe-2017}. 

\begin{figure}[tbph]
\center
\includegraphics*[width=150mm]{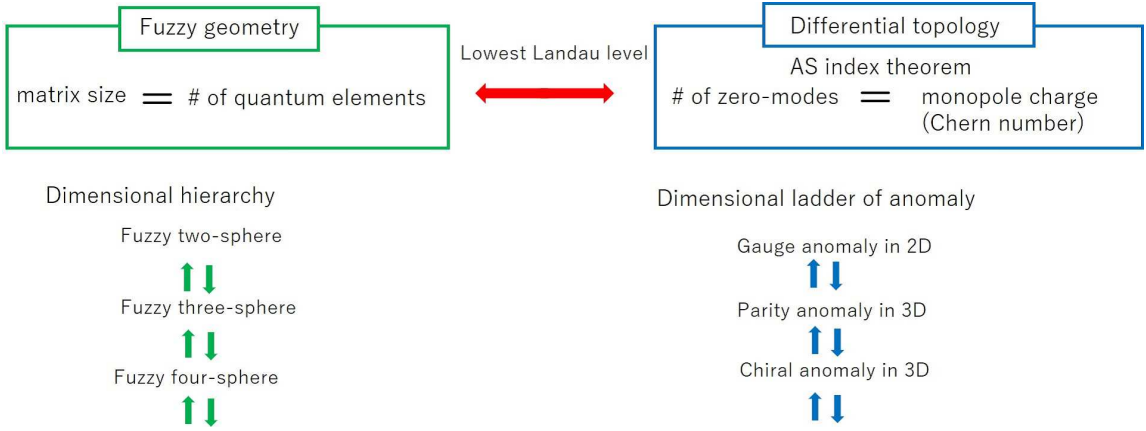}
\caption{ The correspondence between the fuzzy geometry and the differential topology.   
} 
\label{dimr.fig}
\end{figure}

The dimensional hierarchy finds its origin in the differential topology.  In the lowest Landau level, the degenerate  eigenstates constituting the fuzzy spheres are equal to  the zero-modes of the Dirac-Landau operator of the relativistic Landau models.  Meanwhile, the Atiyah-Singer index theorem signifies the equality between  the number of the zero-modes and the Chern number which is the monopole charge in the present Landau models \cite{Dolan2003, Hasebe-2014-1}. 
This implies a close relationship between the non-commutative geometry and the differential topology, since the quantum space of the fuzzy space are spanned by the zero-modes while  the mathematics of the monopole gauge fields is accounted for by the fibre-bundle theory of the differential topology  \cite{Hasebe-2017, Hasebe-2014-1}.   The similar hierarchical structure in the fuzzy geometry and the monopole gauge field is a consequence of this observation.  
In quantum field theory,  
the chiral anomaly  is a  manifestation of  the Atiyah-Singer index theorem and   exhibits a hierarchical structure  
referred to as the dimensional ladder. 
 The dimensional hierarchy is said to be the  Landau model counterpart of the dimensional ladder of quantum anomaly  [Fig.\ref{dimr.fig}]. 

\begin{figure}[tbph]
\center
\includegraphics*[width=150mm]{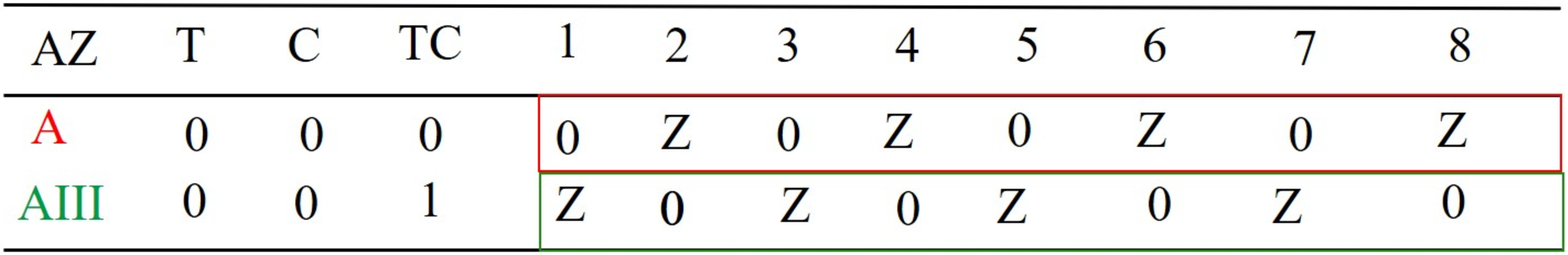}
\caption{The topological table for the A class and the AIII class topological insulators \cite{Ryu-S-F-L-2010, Kitaev2008}. 
} 
\label{table.fig}
\end{figure}

 Incidentally, the even D Landau models correspond to the A-class topological insulators and the odd D Landau models  the AIII-class topological insulators [Fig.\ref{table.fig}]. 
In the topological table, 
it is known that there exists a dimensional relation \cite{Ryu-S-F-L-2010, Kitaev2008}, which  also supports the  idea of the dimensional hierarchy of the Landau models.     
 
\section{Summary}

We discussed the Landau model realization of the fuzzy spheres through the concrete evaluation of the matrix coordinates.   
For the 2D Landau model, the matrix geometries  were identified as the fuzzy spheres in any  Landau levels.   Similarly for the 4D Landau model,  the lowest  Landau level geometry is shown to be the fuzzy four-sphere geometry.  Meanwhile, the higher Landau level geometry turned out to be  the nested fuzzy structure with no classical counterpart.  In the 3D Landau model,   the fuzzy three-sphere geometry was realized at the lowest energy level, and  the hidden one dimension higher fuzzy geometry  was unveiled  in the lowest Landau level. The dimensional hierarchy among the Landau models  is  the Landau model counterpart of the dimensional ladder of quantum anomaly and implies  an intimate relationship between the non-commutative geometry and the differential topology.    



\end{document}